\documentclass[12pt,letterpaper]{article}                  %% LaTeX 2e
\usepackage{osajnl}
\usepackage{overcite,hyperref}

\begin{document}

\title{Probability Density Function of \\ Kerr Effect Phase Noise}

\author{Keang-Po Ho}

\affiliation{StrataLight Communications, Campbell, CA 95008}
\email{kpho@stratalight.com}

\begin{abstract}
The probability density function of Kerr effect phase noise, often called the Gordon-Mollenauer effect, is derived analytically. 
The Kerr effect phase noise can be accurately modeled as the summation of a Gaussian random variable and a noncentral chi-square random variable with two degrees of freedom.
Using the received intensity to correct for the phase noise, the residual Kerr effect phase noise can be modeled as the summation of a Gaussian random variable and the difference of two noncentral chi-square random variables with two degrees of freedom.
The residual phase noise can be approximated by Gaussian distribution better than the Kerr effect phase noise without correction.
\end{abstract}
% REPLACE WITH CORRECT OCIS CODES FOR YOUR ARTICLE
                          % NOTE: \ocis{} IS ALIASED TO \pacs{} BUT MUST
                          % FORMAT THE TERMS CORRECTLY FOR EACH JOURNAL

\maketitle %% NULL FUNCTION WITH LATEX 2e

\section{Introduction}

Gordon and Mollenauer \cite{gordon} showed that when optical amplifiers are used to compensate for fiber loss, the interaction of amplifier noise and the Kerr effect causes phase noise, often called the Gordon-Mollenauer effect or nonlinear phase noise.
By broadening the signal linewidth \cite{ryu}, Kerr effect phase noise degrades both phase-shifted keying (PSK) and differential phase-shift keying (DPSK) systems that have renewed attention recently  \cite{gnauck, griffin, zhu}. 
Because the Kerr effect phase noise is correlated with the received intensity, the received intensity can be used to correct the Kerr effect phase noise \cite{liu, xu, ho}.
The transmission distance can be doubled if the Kerr effect phase noise is the dominant impairment \cite{liu, ho}.

Usually, the performance of the system is estimated based on the variance of the Kerr effect phase noise \cite{gordon, liu, xu, ho}. 
The probability density function (p.d.f.) is required to better understand the system and evaluates the system performance. 
This paper provides an analytical expression of the p.d.f. for the Kerr effect phase noise with \cite{liu, xu, ho} and without \cite{gordon} the correction by the received intensity.
The characteristic functions are first derived analytically and the p.d.f.'s are the inverse Fourier transform of the characteristic functions.

\section{Probability Density Function}

For simplicity and without loss of generality, assume that the total Kerr effect phase noise is \cite{gordon, liu, ho}

\begin{equation}
\phi_{\mathrm{LN}}  = |A + n_1|^2 + |A + n_1 + n_2|^2 + \cdots + |A + n_1 + \cdots + n_N|^2,
\label{phinl}
\end{equation}

\noindent where $A$ is a real number representing the amplitude of the transmitted signal, $n_k = x_k +i y_k$, $k = 1, \ldots, N$, are the optical amplifier noise introduced into the system at the $k$th fiber span, $n_k$ are independent identically distributed (i.i.d.) complex zero-mean circular Gaussian random variables with $E\{x_k^2\} = E\{y_k^2\} = E\{|n_k|^2\}/2 = \sigma^2$, where $\sigma^2$ is the noise variance per dimension per span.  
The product of fiber nonlinear coefficient and the effective length per span $\gamma L_{\mathrm{eff}}$ is
ignored in (\ref{phinl}) for simplicity \cite{gordon, liu, ho}. 

First, we consider the random variable of

\begin{equation}
\varphi_1 = |A + x_1|^2 + |A + x_1 + x_2|^2 + \cdots + |A + x_1 + \cdots + x_N|^2.
\label{varphi}
\end{equation}

\noindent The overall Kerr effect phase noise (\ref{phinl}) is $\phi_{\mathrm{LN}} = \varphi_1 + \varphi_2$, where 

\begin{equation}
\varphi_2 = y_1^2 + |y_1 + y_2|^2 + \cdots + |y_1 + \cdots + y_N|^2
\label{varphi2}
\end{equation}

\noindent is independent of $\varphi_1$ and has a p.d.f. equal to that of $\varphi_1$ when $A = 0$. The random variable of (\ref{varphi}) can be expressed as

\begin{equation}
\varphi_1 = NA^2 + 2A \vec{w}^T \vec{x} + \vec{x}^T {\mathcal C} \vec{x},
\label{var1}
\end{equation}

\noindent where $\vec{w} = \{N, N-1, \ldots, 2, 1\}^T$, $\vec{x} = \{x_1, x_2, \ldots , x_N\}^T$, and the covariance matrix ${\mathcal C} = {\mathcal M}^T {\mathcal M}$ with

\begin{equation}
{\mathcal M} = \left[ 
\begin{array}{ccccc}
1 & 0 & 0 & \cdots & 0 \\
1 & 1 & 0 & \cdots & 0 \\
1 & 1 & 1 & \cdots & 0 \\
\vdots & \vdots & \vdots & \ddots & \vdots \\
1 & 1 & 1 & \cdots & 1
\end{array} 
\right].
\end{equation}

The p.d.f. of $\vec{x}$ is $ {(2 \pi \sigma^2)^{-\frac{N}{2}}}\exp \left( - { \vec{x}^T \vec{x}}/{2 \sigma^2} \right)$. 
The characteristic function of $\varphi_1$, $\Psi_{\varphi_1}(\nu) = E\left\{ \exp \left( j \nu \varphi_1 \right) \right\}$, is

\begin{equation}
\Psi_{\varphi_1}(\nu) =\frac{\exp(j \nu N A^2) } {(2 \pi \sigma^2)^{\frac{N}{2}}}
			\int \exp \left[ 2 j \nu A \vec{w}^T \vec{x} -
		\vec{x}^T \Gamma \vec{x} \right] \mathrm{d} \vec{x},
\label{cf1}
\end{equation}

\noindent where $\Gamma = {{\mathcal I}/(2 \sigma^2) - j \nu  {\mathcal C }}$ and ${\mathcal I}$ is an $N \times N$ identity matrix. 
Using the relationship of 

\begin{equation}
 \vec{x}^T \Gamma \vec{x} - 
2 j \nu A \vec{w}^T \vec{x}  = \left( \vec{x} - j \nu A \Gamma^{-1} \vec{w} \right)^T \Gamma \left( \vec{x} - j \nu A \Gamma^{-1} \vec{w} \right) 
+ \nu^2 A^2 \vec{w}^T \Gamma^{-1} \vec{w},
\end{equation}

\noindent with some algebra, the characteristic function (\ref{cf1}) is

\begin{equation}
\Psi_{\varphi_1}(\nu)
=   \frac{\exp{\left[ j \nu N A^2-\nu^2 A^2 \vec{w}^T \Gamma^{-1} \vec{w} \right]}}
  { {(2 \sigma^2)^{\frac{N}{2}}} { \det[\Gamma]^{1/2}}}, 
\label{cf1a}
\end{equation}

\noindent where $\det[\cdot]$ is the determinant of a matrix. The characteristic function (\ref{cf1a}) is

\begin{equation}
\Psi_{\varphi_1}(\nu) = 
	\frac{\exp \left[ j \nu N A^2 - 2 \sigma^2 \nu^2 A^2 \vec{w}^T ({{\mathcal I} - 2 j \nu \sigma^2 {\mathcal C }})^{-1} \vec{w} \right]}
	{\det \left[ {\mathcal I} - 2 j \nu \sigma^2 {\mathcal C }\right]^{\frac{1}{2}}}.
\label{cf2}
\end{equation}

\noindent Substitute $A = 0$ into (\ref{cf2}), the characteristic function of $\varphi_2$  is $\Psi_{\varphi_2}(\nu) = {\det \left[ {\mathcal I} - 2 j \nu \sigma^2 {\mathcal C }\right]^{-\frac{1}{2}}}$. The characteristic function of $\phi_{\mathrm{LN}}$ is $\Psi_{\phi_{\mathrm{LN}}}(\nu)  = \Psi_{\varphi_1}(\nu) \Psi_{\varphi_2}(\nu) $, or 

\begin{equation}
\Psi_{\phi_{\mathrm{LN}}}(\nu) = 
	\frac{\exp \left[ j \nu N A^2 - 2 \sigma^2 \nu^2 A^2 \vec{w}^T ({{\mathcal I} - 2 j \nu \sigma^2 {\mathcal C }})^{-1} \vec{w} \right]}
	{\det \left[ {\mathcal I} - 2 j \nu \sigma^2 {\mathcal C }\right]}.
\label{cfnl}
\end{equation}

If the covariance matrix ${\mathcal C }$ has eigenvalues and eigenvectors of $\lambda_k$, $\vec{v}_k$, $k = 1, 2, \ldots, N$, respectively, the characteristic function (\ref{cfnl}) becomes

\begin{equation}
\Psi_{\phi_{\mathrm{LN}}}(\nu) = \frac{ \displaystyle \exp \left[ 
	j \nu N A^2 - 2 \sigma^2 \nu^2 A^2 \sum_{k=1}^N \frac{(\vec{v}_k^T \vec{w})^2}{1 - 2 j \nu \sigma^2 \lambda_k}
	\right] }
{\displaystyle \prod_{k = 1}^{N} \left(1 - 2 j \nu \sigma^2 \lambda_k \right) },
\label{oPsiEig}
\end{equation}

\noindent and can be rewritten to 

\begin{equation}
\displaystyle \Psi_{\phi_{\mathrm{LN}}}(\nu)  = \prod_{k = 1}^N \frac{1}{1 - 2 j \nu \sigma^2 \lambda_k} 
	\exp \left[ \frac {j \nu A^2 (\vec{v}_k^T \vec{w})^2/\lambda_k} {1 - 2 j \nu \sigma^2 \lambda_k} \right].
\label{oPsiEig1}
\end{equation}

\noindent From the characteristic function (\ref{oPsiEig1}), the random variable of $\phi_{\mathrm{LN}}$ (\ref{phinl}) is the summation of $N$ independently distributed noncentral $\chi$-square random variables with two degrees of freedom \cite{proakis}. 
Without going into detail, the matrix 

\begin{equation}
{\mathcal C}^{-1} = \left[
\begin{array}{cccccc}
1 & -1 & 0 & \cdots &0 &  0 \\
-1 & 2 & -1 & \cdots & 0 & 0 \\
0 & -1 & 2 & \cdots & 0 & 0 \\
\vdots & \vdots & \vdots & \ddots & \vdots & \vdots \\
0 & 0 & 0 & \cdots & -1 & 2
\end{array} 
\right]
\label{invM}
\end{equation}

\noindent is approximately a Toepliz matrix for the series of
$\ldots, 0, -1, 2, -1, 0, \ldots$ For large $N$, the eigenvalues of the covariance matrix of ${\mathcal C}$ are asymptotically equal to \cite{gray}

\begin{equation}
\frac{1}{\lambda_k}  \approx  2  \left[ 1 - \cos \left( \frac{(2 k + 1) \pi}{2 N} \right) \right] 
	=   4 \sin^2 \left( \frac{(2 k - 1) \pi}{4 N} \right) , \hspace{0.25cm} k = 1, \ldots, N.
\label{eig}
\end{equation} 

\noindent The values of (\ref{eig}) are the discrete Fourier transform of each row of the matrix ${\mathcal C}^{-1}$. 
The eigenvalues of the covariance matrix of $\mathcal C$ are all positive and multiple to unity. 

With the correction of phase noise using received intensity \cite{liu, xu, ho}, the residual nonlinear phase noise is

\begin{eqnarray}
\phi_{\mathrm{RES}} & = & |A+n_1|^2 + |A+ n_1+n_2|^2 +  \cdots  + |A+ n_1 + \cdots +n_{N-1}|^2 \nonumber \\
	& & - (\alpha_{\mathrm{opt}}-1) |A+ n_1 + \cdots + n_{N}|^2,
\label{phires}
\end{eqnarray}

\noindent As from the Appendix, $\alpha_{\mathrm{opt}} \approx (N+1)/2$ is the optimal scale factor to correct the Kerr effect phase noise (\ref{phinl}) using the received intensity of $|A+ n_1 + \cdots + n_{N}|^2$. 
The random variable corresponding to $\varphi_1$ (\ref{var1}) becomes

\begin{equation}
(N-\alpha_{\mathrm{opt}})A^2 + 2A \vec{w_r}^T \vec{x} + \vec{x}^T {\mathcal C_r} \vec{x},
\end{equation}

\noindent where $\vec{w_r} =\vec{w}- \alpha_{\mathrm{opt}} \cdot \{1, 1, \ldots, 1\}^{T}$ and

\begin{equation}
{\mathcal C}_r = ({\mathcal M} - {\mathcal L})^{T}({\mathcal M} - {\mathcal L}) 
	- (\alpha_{\mathrm{opt}}-1){\mathcal L}^{T}{\mathcal L},
\end{equation}

\noindent where 

\begin{equation}
{\mathcal L} = \left[
\begin{array}{ccccc}
	0 & 0 & \cdots & 0 & 0	\\
	\vdots & \vdots & \ddots & \vdots & \vdots \\
	0 & 0 & \cdots & 0 & 0 \\
	1 & 1 & \cdots & 1 & 1
\end{array} 
\right].
\end{equation} 

Following the procedure from (\ref{var1}) to (\ref{cfnl}), the characteristic function of $\phi_{\mathrm{RES}}$ is

\begin{equation}
\Psi_{\phi_{\mathrm{RES}}}(\nu) =  
	 \frac{\exp \left[ -j \nu (N - \alpha_{\mathrm{opt}}) A^2- 2 \sigma^2 \nu^2 A^2 \vec{w_r}^T ({{\mathcal I} - 2 j \nu 	\sigma^2 {\mathcal C}_r})^{-1} \vec{w_r} \right]}
		{ \det \left[ {\mathcal I} - 2 j \nu \sigma^2 {\mathcal C}_r\right]}.
\label{cfres}
\end{equation}

\noindent The characteristic functions of $\phi_{\mathrm{RES}}$ in the form of eigenvalues and eigenvectors are similar to that of (\ref{oPsiEig}) and (\ref{oPsiEig1}). 
The characteristic functions of $\phi_{\mathrm{RES}}$ has the same expression as (\ref{oPsiEig1}) using a new set of eigenvalues and eigenvectors of the covariance matrix ${\mathcal C}_r$ and the vector of $\vec{w_r}$.

Except the first and last rows, the matrix ${\mathcal C}_r^{-1}$ is also approximately a Toepliz matrix for the series of $\ldots, 0, -1, 2, -1, 0, \ldots$ For large $N$, the eigenvalues of ${\mathcal C_r}$ are asymptotically equal to 

\begin{equation}
\frac{1}{\lambda_k}  \approx   
	 4 \sin^2 \left[ \frac{(k - 1.25) \pi}{2 (N-1)} \right] , \qquad k = 2, \ldots, N, \qquad
\lambda_1 \approx -\sum_{k=2}^{N} \lambda_k.
\label{eig1}
\end{equation} 

\noindent Other than the largest one in absolute value, the eigenvalues of ${\mathcal C}_r$ are all positive. 
All eigenvalues of the covariance ${\mathcal C}_r$ sum to approximately zero and multiple to  $\alpha_{\mathrm{opt}} -1 \approx (N-1)/2$.

\section{Numerical Results and Random Variable Models}

The p.d.f.'s of both $\phi_{\mathrm{NL}}$ (\ref{phinl}) and $\phi_{\mathrm{RES}}$ (\ref{phires}) can be calculated by taking the inverse Fourier transform of the corresponding characteristic functions of $\Psi_{\phi_{\mathrm{NL}}}(\nu)$ (\ref{cfnl}) and $\Psi{\phi_{\mathrm{RES}}}(\nu)$ (\ref{cfres}), respectively. 
Fig. \ref{figpdf} shows the p.d.f. of $\phi_{\mathrm{NL}}$ (\ref{phinl}) and $\phi_{\mathrm{RES}}$ (\ref{phires}). 
Fig. \ref{figpdf} is plotted for the case that the optical signal-to-noise ratio $\rho_O = A^2/(2 N \sigma^2) = 18$, corresponding to an error probability of $10^{-9}$ if the amplifier noise is the only impairment. 
The number of span is $N = 32$. 
The $x$-axis is normalized with respect to $NA^2$, approximately equal to the mean Kerr effect phase noise from the Appendix. 

Fig. \ref{figpdf} can confirm that using the received intensity to correct for Kerr effect phase noise, the standard deviation of Kerr effect phase noise can be reduced by a factor of two \cite{liu, xu, ho}. 
The Appendix shows that the variance of nonlinear phase noise can be reduced by approximately a factor of four.

From the characteristic function of (\ref{oPsiEig1}), the random variables of  both $\phi_{\mathrm{NL}}$ (\ref{phinl}) and $\phi_{\mathrm{RES}}$ (\ref{phires}) can be modeled  as the combination of $N = 32$ independently distributed noncentral $\chi$-square random variables with two degrees of freedom.
Some studies \cite{gordon, liu, xu} implicitly assume a Gaussian distribution by using the $Q$-factor to characterize the random variables. 
When many independently distributed random variables with more or less the same variance are summed (or subtracted) together, the summed random variable approaches the Gaussian distribution.
For the characteristic function of (\ref{oPsiEig1}), the Gaussian assumption is valid only if the eigenvalues $\lambda_k$ are more or less the same. 
From (\ref{eig}), the largest eigenvalue $\lambda_1$ of the covariance matrix $\mathcal C$ is about nine times larger than the second largest eigenvalue $\lambda_2$. 
From (\ref{eig1}), the two largest eigenvalues $\lambda_1$ and $\lambda_2$ of the covariance matrix ${\mathcal C}_r$ are about $5.5$ times larger than the third largest eigenvalue $\lambda_3$.
The approximation of (\ref{eig}) is accurate within 3.2\% for $N = 32$.
The approximation of (\ref{eig1}) is not as good as that for (\ref{eig}) and accurate within 10\% for $N = 32$.

While the Gaussian assumption for both $\phi_{\mathrm{NL}}$ (\ref{phinl}) and $\phi_{\mathrm{RES}}$ (\ref{phires}) may not be valid, other than the noncentral $\chi$-square random variables with two degrees of freedom corresponds to some large eigenvalues, the other random variables should sum to Gaussian distribution.
By modeling the summation of random variables with smaller eigenvalues as Gaussian distribution, the nonlinear phase noise of (\ref{oPsiEig1}) can be modeled as a summation of two or three instead of $N = 32$ independently distributed random variables. 

Note that the variance of the noncentral $\chi$-square random variables with two degrees of freedom in (\ref{oPsiEig1}) is $4\sigma^4 \lambda_k^2 + 4A^2 (\vec{v}_k^T \vec{w})^2$ \cite{proakis}. 
While the above reasoning just takes into account the contribution from the eigenvalue of $\lambda_k$ but ignores the contribution from the eigenvector $\vec{v}_k$, numerical results show that the variance of each individual noncentral $\chi$-square random variable increases with the corresponding eigenvalue of $\lambda_k$. 
Later part of this paper also validates the argument.

From Fig. \ref{figpdf}, the p.d.f. of $\phi_{\mathrm{NL}}$ has significant difference with that of a Gaussian distribution. 
Fig. \ref{figconv2} divides the p.d.f. of $\phi_{\mathrm{NL}}$ into the convolution of two parts. 
The first part has no observable difference with a Gaussian p.d.f. and corresponds to the second largest to the smallest eigenvalues, $\lambda_k, k = 2, \ldots, N$, of the characteristic function (\ref{oPsiEig1}).
The second part is a noncentral $\chi$-square p.d.f. with two degrees of freedom and corresponds to the largest eigenvalue $\lambda_1$, where $\sigma^2 \lambda_1 \approx 2/(\pi^2 \rho_O) \cdot NA^2$. 
The p.d.f. of $\phi_{\mathrm{NL}}$ in Fig. \ref{figpdf} is also plotted in Fig. \ref{figconv2} for comparison.
The mean and variance of the first part Gaussian random variable are $\sum_{k=2}^{N} A^2 (\vec{v}_k^T \vec{w})^2/\lambda_k + 2\sigma^2 \lambda_k$ and $4\sum_{k=2}^{N}  \sigma^4 \lambda_k^2 + A^2 (\vec{v}_k^T \vec{w})^2$, respectively. 
The second part noncentral $\chi$-square p.d.f. with two degrees of freedom has a variance parameter of $\sigma^2 \lambda_1$ and noncentrality parameter of $A^2 (\vec{v}_1^T \vec{w})^2/\lambda_1$ \cite{proakis}.

To verify that the modeling in Fig. \ref{figconv2} is accurate, the cumulative tail probabilities are calculated by $\int_{-\infty}^{x} p(\xi) d \xi$ and $\int_{x}^{+\infty} p(\xi) d \xi$, where $p(\xi)$ is the p.d.f. 
Fig. \ref{fignl} shows the cumulative tail probabilities as a function of $Q$-factor for $\phi_{\mathrm{NL}}$, defined as $Q = (\phi_{\mathrm{NL}} - \overline{\phi_{\mathrm{NL}}})/\sigma_{\phi_{\mathrm{NL}}}$, where $\overline{\phi_{\mathrm{NL}}}$ and $\sigma_{\phi_{\mathrm{NL}}}^2$ are the mean and variance of the Kerr effect phase noise given in the Appendix.
Using Gaussian approximation, this definition of $Q$-factor gives the same tail probability or error probability \cite{gordon, liu, xu} of $\frac{1}{2} \mathrm{erfc} (|Q| /\sqrt{2})$, where $\mathrm{erfc}(\cdot)$ is the complementary error function.    
Fig. \ref{fignl} shows the cumulative tail probabilities calculated by numerical integration according to (\ref{cfnl}) as circle, the model as the summation of a Gaussian and a noncentral $\chi$-squarerandom variable with two degrees of freedom of Fig. \ref{figconv2} as solid lines, and the Gaussian assumption as dotted lines.
From Fig. \ref{fignl}, the Gaussian approximation by $Q$-factor is not accurate, especially for the tail probability for less than the mean.   
From Fig. \ref{fignl}, the Kerr effect phase noise can be modeled very accurately as the summation of a Gaussian random variable and a noncentral $\chi$-square random variable with two degrees of freedom.
From Fig. \ref{figconv2}, the noncentral $\chi$-square random variable with two degrees of freedom corresponding to $\lambda_1$ has a very large variance such that the p.d.f. of $\phi_{\mathrm{LN}}$ in Fig. \ref{figpdf} has significant difference with a Gaussian p.d.f.

Instead of the combination of $N = 32$ noncentral $\chi$-square random variables with two degrees of freedom, similar to the decomposition of Fig. \ref{figconv2}, the random variable of $\phi_{\mathrm{RES}}$ can be modeled as the summation of a Gaussian random variable and the difference of two noncentral $\chi$-square random variables with two degrees of freedom. 
Fig. \ref{figconv3} shows that the p.d.f. of  $\phi_{\mathrm{RES}}$ as the convolution of a Gaussian p.d.f. and two noncentral $\chi$-square p.d.f.'s with two degrees of freedom.
The two noncentral $\chi$-square random variables correspond to the two largest eigenvalues of the covariance matrix ${\mathcal C}_r$ with more or less the same magnitude but different signs.
The Gaussian random variable corresponds to the summation of $N-2$  noncentral $\chi$-square random variables with two degrees of freedom for the eigenvalues of $\lambda_3, \dots, \lambda_N$.
Because the variance parameter of $\sigma^2 \lambda_1$ is negative, the corresponding random variable in (\ref{oPsiEig1}) is the negative of a noncentral $\chi$-square random variable with two degrees of freedom. 
The p.d.f. corresponding to $\lambda_1$ in Fig. \ref{figconv3} is the mirror image of a noncentral $\chi$-square p.d.f. with two degrees of freedom with respect to the $y$-axis.
The random variable corresponding to the combined term of both $\lambda_1$ and $\lambda_2$ in (\ref{oPsiEig1}) is the difference of two noncentral $\chi$-square random variables with two degrees of freedom.   

Fig. \ref{figres} shows the cumulative tail probabilities as a function of $Q$-factor for $\phi_{\mathrm{RES}}$, defined as $Q = (\phi_{\mathrm{RES}} - \overline{\phi_{\mathrm{RES}}})/\sigma_{\phi_{\mathrm{RES}}}$, where $\overline{\phi_{\mathrm{RES}}}$ and $\sigma_{\phi_{\mathrm{RES}}}^2$ are the mean and variance of the residual phase noise shown in the Appendix.   
The cumulative tail probabilities calculated by numerical integration according to (\ref{cfres}) is shown as circle, the model as the summation of a Gaussian random variable and the difference of two  noncentral $\chi$-square random variables with two degrees of freedom of Fig. \ref{figconv3} is shown as solid lines, and the Gaussian assumption \cite{gordon, liu, xu} is shown as dotted lines.
From Figs. \ref{figpdf} and \ref{figconv3}, the p.d.f. of $\phi_{\mathrm{RES}}$ resembles a Gaussian p.d.f. with mean and variance from \citeleft\citenum{ho}\citeright  and the Appendix. 
The residual Kerr effect phase noise of $\phi_{\mathrm{RES}}$ can be modeled accurately as a Gaussian random variable, especially for the tail probabilities less than the mean.
Even for the tail probabilities larger than the mean, the Gaussian model for $\phi_{\mathrm{RES}}$ is better than that for $\phi_{\mathrm{NL}}$. 
If the tail probabilities for above $10^{-5}$ is for interests, Gaussian approximation for $\phi_{\mathrm{RES}}$ can be used.

\section{Conclusion}

The characteristic functions of Kerr effect phase noise, with and without the correction using the received  intensity, are derived analytically as product of $N$ noncentral $\chi$-square characteristic functions with two degrees of freedom. 
The p.d.f.'s are calculated exactly as the inverse Fourier transform of the characteristic functions.
The p.d.f. of the Kerr effect phase noise can be modeled as the convolution of a Gaussian p.d.f. and a noncentral $\chi$-square p.d.f. with two degrees of freedom. 
Using the received intensity to correct for the phase noise, the p.d.f. of the residual Kerr effect phase noise can be modeled accurately as the convolution of a Gaussian p.d.f and two noncentral $\chi$-square p.d.f.'s with two degrees of freedom.
The Gaussian approximation of the residual Kerr effect phase noise is much better than that for Kerr effect phase noise.

\section*{Appendix: Optimal Linear Compensator}

This appendix shows important results from \citeleft\citenum{ho}\citeright. 
The optimal scale factor to minimimize the variance of $\phi_{\mathrm{RES}}$ is

\begin{equation}
\alpha_{\mathrm{opt}} = \frac{N+1}{2} \cdot \frac{A^2 +(2N+1)\sigma^2/3}{A^2 + N\sigma^2}
	\approx \frac{N+1}{2}.
\end{equation}

The variance of the residual nonlinear phase noise of (\ref{phires}) is reduced to

\begin{equation}
\sigma^2_{\phi_{\mathrm{RES}}} = (N-1)N(N+1)\sigma^2 \cdot 
	\frac{A^4 + 2N \sigma^2 A^2 + (2N^2+1)\sigma^4/3}{3(A^2+ N \sigma^2)}
\end{equation}

\noindent from that of the Kerr effect phase noise of

\begin{equation}
\sigma^2_{\phi_{\mathrm{NL}}} = \frac{4}{3} N(N+1) \sigma^2 
	\left[ ( N + \frac{1}{2}) A^2 + (N^2 + N + 1) \sigma^2 \right]
\end{equation}  

The mean of the Kerr effect phase noise (\ref{phinl}) is

\begin{equation}
\overline{\phi_{\mathrm{NL}}} = N \left[ A^2 + (N+1) \sigma^2 \right]
\end{equation}

The mean of the residual nonlinear phase noise is

\begin{equation}
\overline{\phi_{\mathrm{RES}}} = \overline{\phi_{\mathrm{NL}}} - 
	\alpha_{\mathrm{opt}} (A^2 + 2N \sigma^2)
\end{equation}

\newpage

\section*{List of Figure Captions}

\begin{figure}[h]\centerline{\scalebox{0.75}{\includegraphics{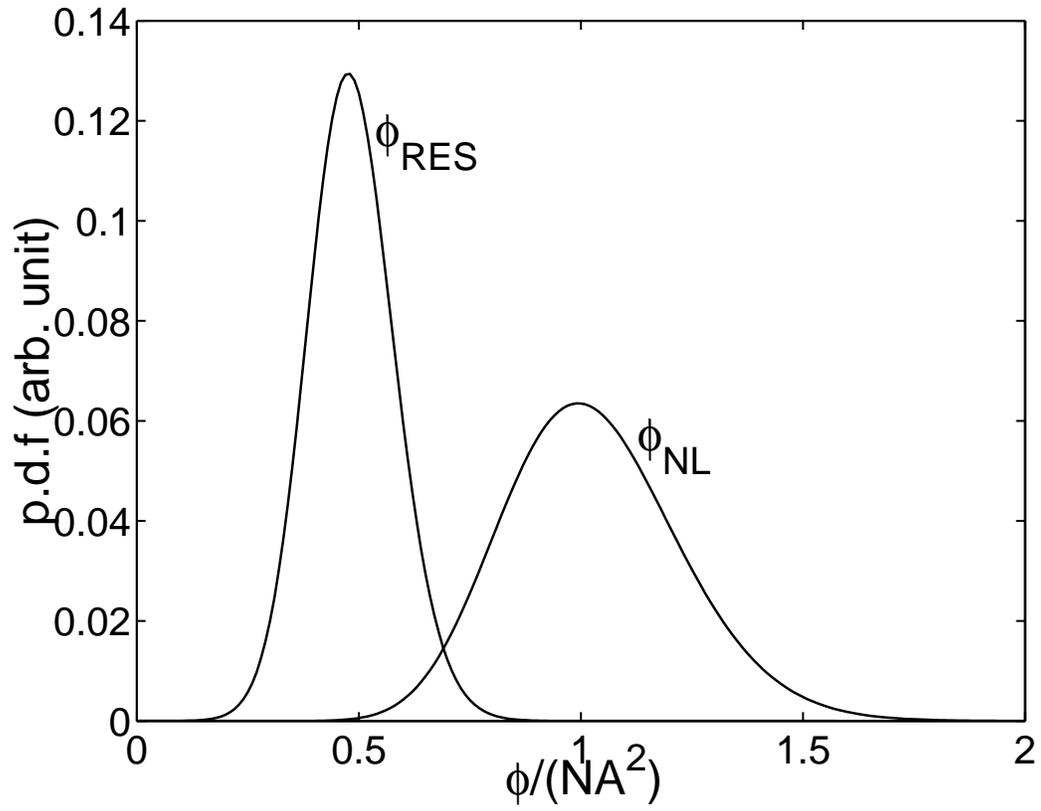}}}
\caption{The p.d.f. of both $\phi_{\mathrm{NL}}$ and $\phi_{\mathrm{RES}}$.}
\label{figpdf}
\end{figure}

Fig. \ref{figpdf}. The p.d.f. of both $\phi_{\mathrm{NL}}$ and $\phi_{\mathrm{RES}}$.

\begin{figure}[h]\centerline{\scalebox{0.75}{\includegraphics{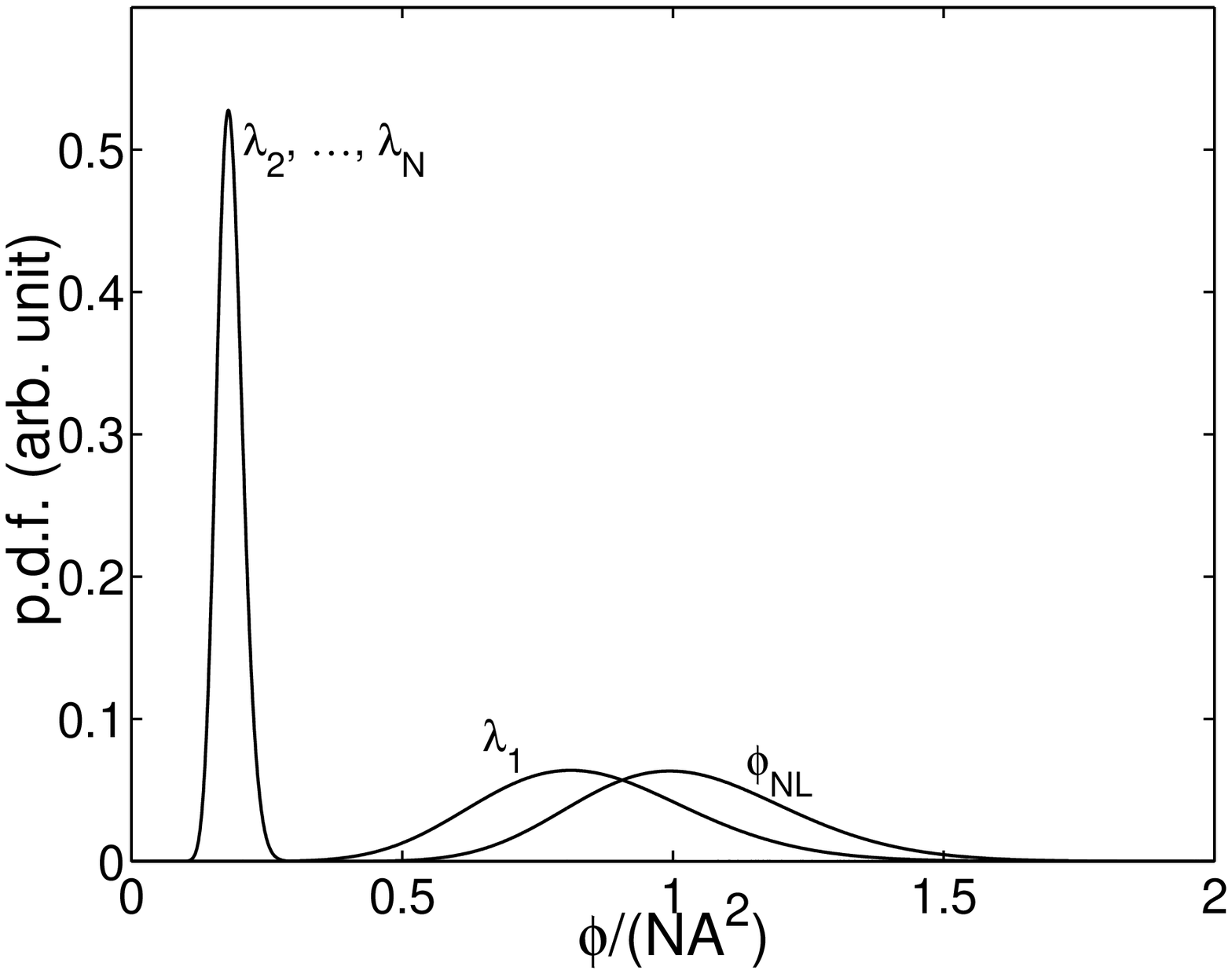}}}
\caption{The p.d.f. of $\phi_{\mathrm{NL}}$ is the convolution of a Gaussian p.d.f. and a noncentral $\chi$-square p.d.f. with two degrees of freedom.}
\label{figconv2}
\end{figure}

\noindent Fig. \ref{figconv2}. The p.d.f. of $\phi_{\mathrm{NL}}$ is the convolution of a Gaussian p.d.f. and a noncentral $\chi$-square p.d.f. with two degrees of freedom.

\begin{figure}[h]\centerline{\scalebox{0.75}{\includegraphics{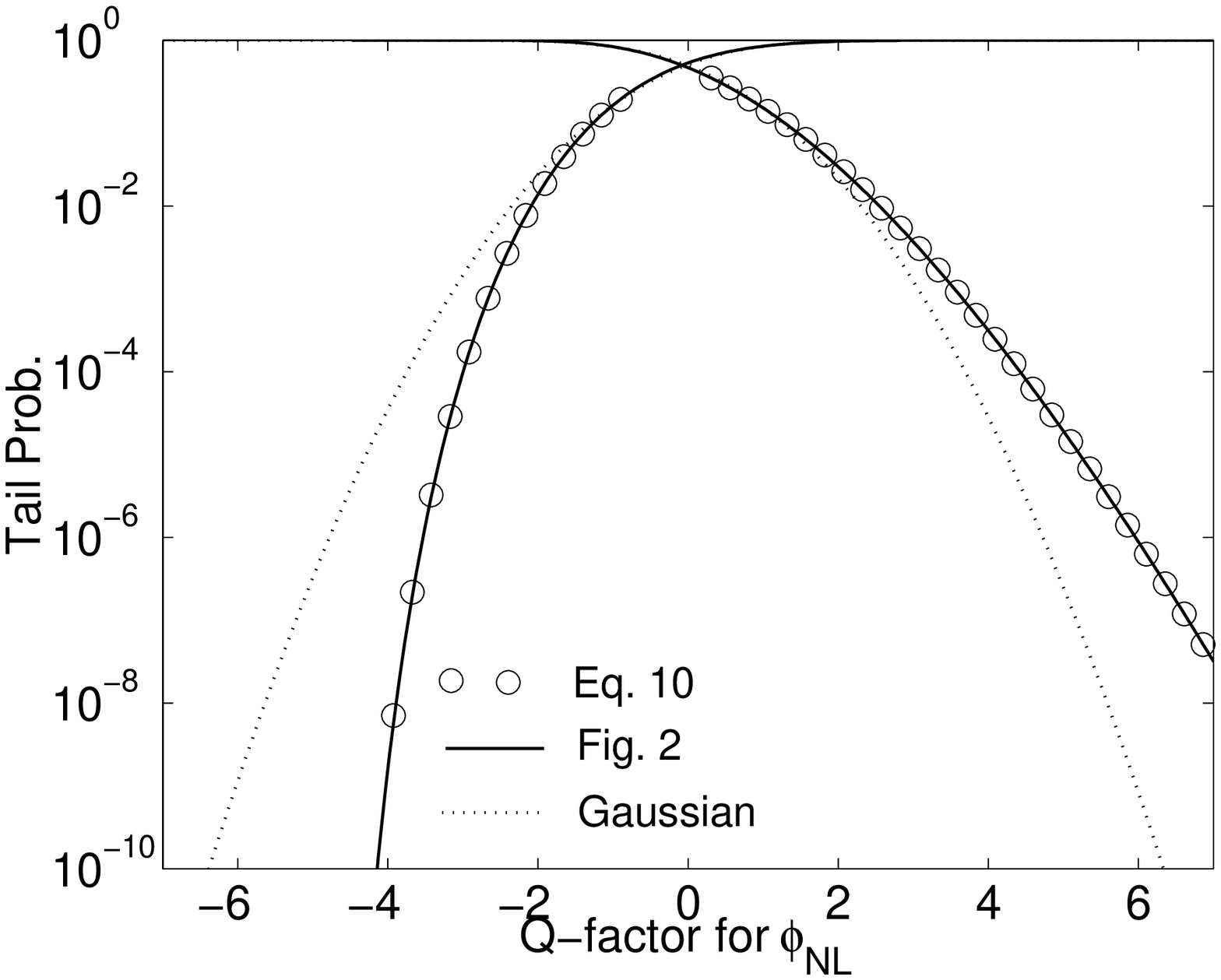}}}
\caption{The cumulative tail probability of $\phi_{\mathrm{NL}}$ as compared with the model of Fig. \ref{figconv2} and Gaussian approximation.}
\label{fignl}
\end{figure}

\noindent Fig. \ref{fignl}. The cumulative tail probability of $\phi_{\mathrm{NL}}$ as compared with the model of Fig. \ref{figconv2} and Gaussian approximation.

\begin{figure}[h]\centerline{\scalebox{0.75}{\includegraphics{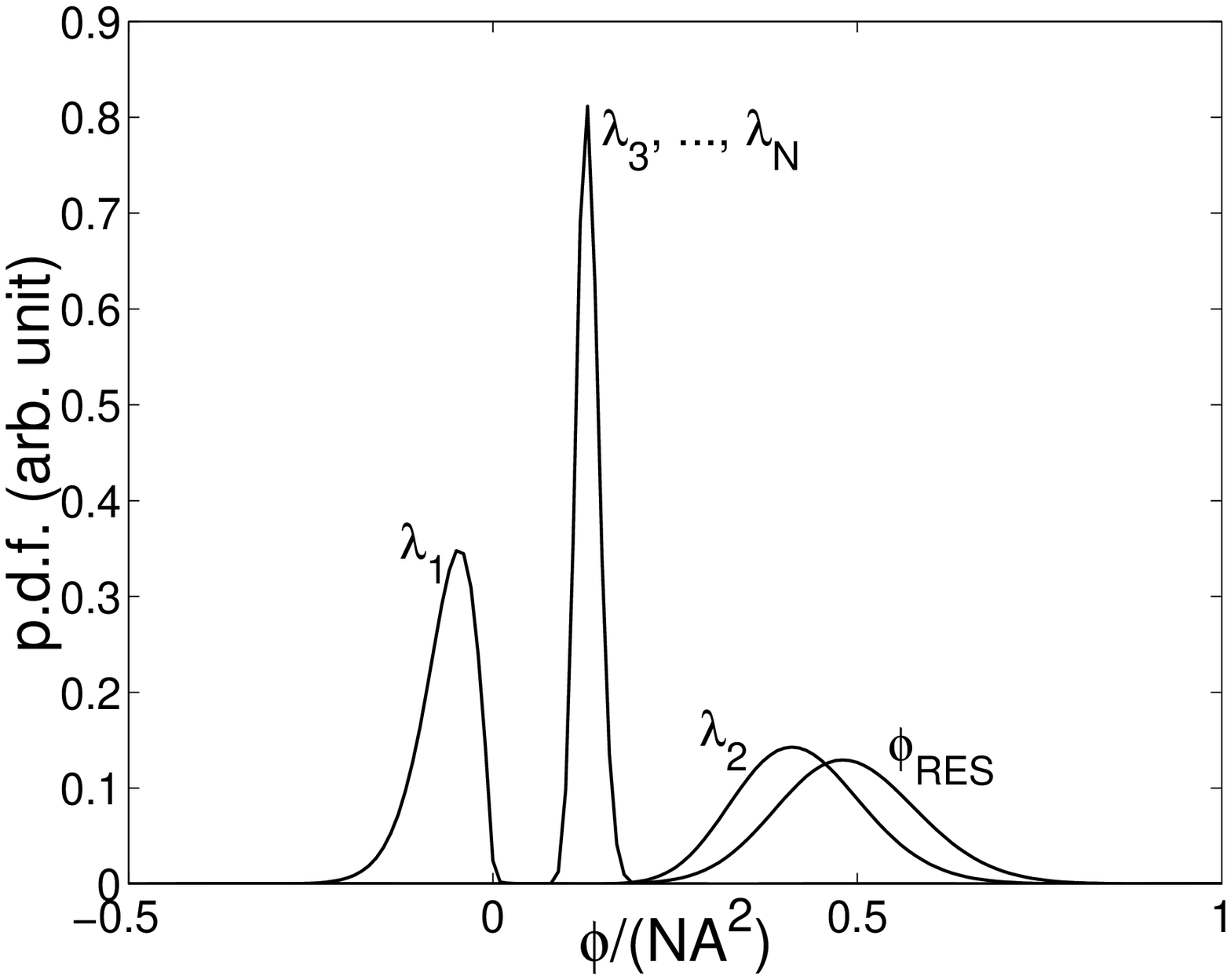}}}
\caption{The p.d.f. of $\phi_{\mathrm{RES}}$ is the convolution of a Gaussian p.d.f. and two noncentral $\chi$-square p.d.f.'s with two degrees of freedom.}
\label{figconv3}
\end{figure}

\noindent Fig. \ref{figconv3}. The p.d.f. of $\phi_{\mathrm{RES}}$ is the convolution of a Gaussian p.d.f. and two noncentral $\chi$-square p.d.f.'s with two degrees of freedom.

\begin{figure}[h]\centerline{\scalebox{0.75}{\includegraphics{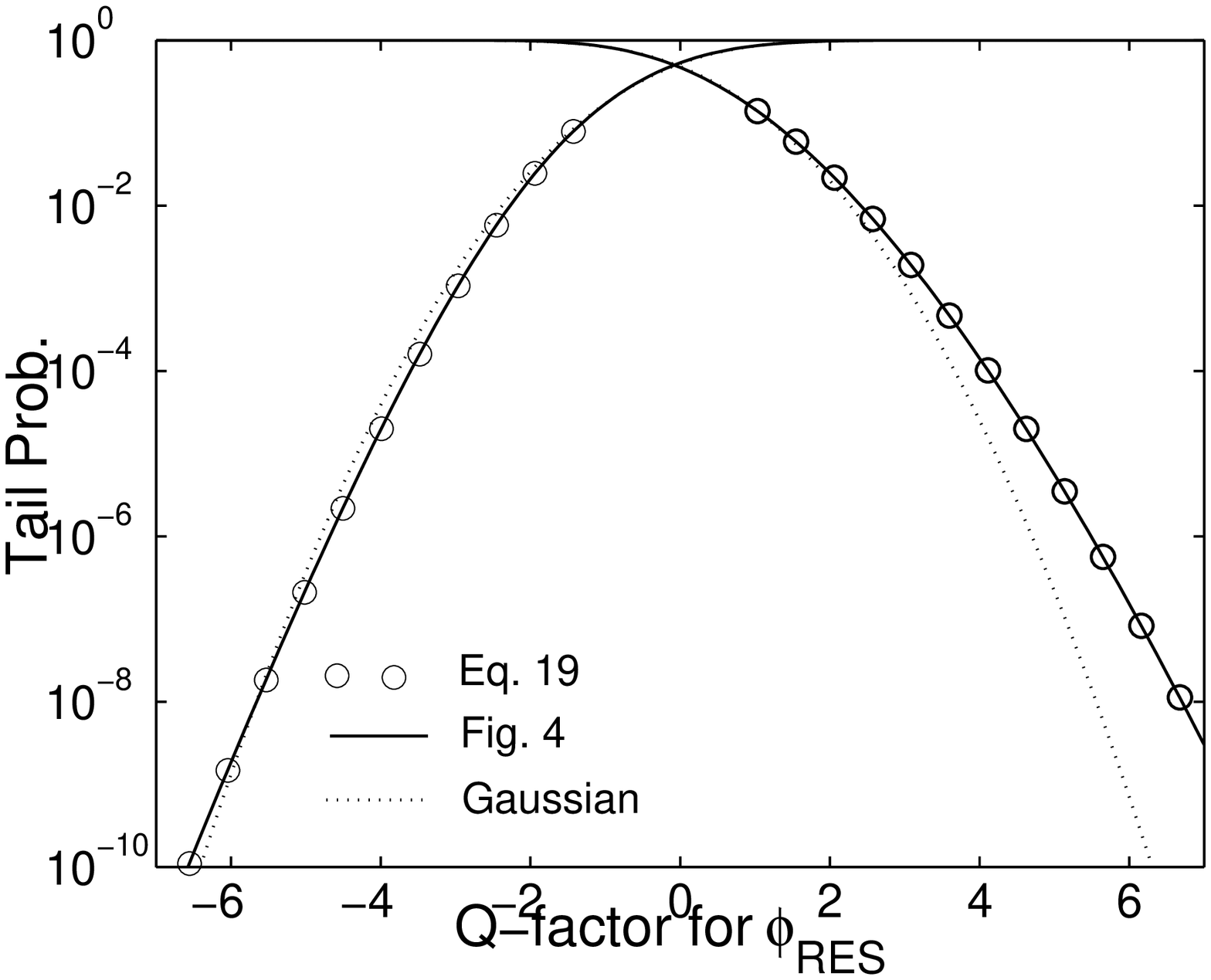}}}
\caption{The cumulative tail probability of $\phi_{\mathrm{RES}}$ as compared with the model of Fig. \ref{figconv3} and Gaussian approximation.}
\label{figres}
\end{figure}

\noindent Fig. \ref{figres}. The cumulative tail probability of $\phi_{\mathrm{RES}}$ as compared with the model of Fig. \ref{figconv3} and Gaussian approximation.

\newpage

\end{document}